\documentclass[bimj,fleqn]{w-art}
\usepackage{times}
\usepackage{w-thm}
\usepackage[authoryear]{natbib}

\usepackage{mathtools}
\usepackage{graphicx}
\usepackage{xcolor}
\usepackage[mathscr]{euscript}
\usepackage{color}
\usepackage{colortbl}
\usepackage{tikz}
\usepackage{pgfplots}
\usepackage{booktabs}
\usepackage{float}

\theoremstyle{plain}

\theoremstyle{definition}

\usepackage[]{graphicx}
\chardef\bslash=`\\ 

\begin{document}

\newcommand{\BLAsolid}{\raisebox{2pt}{\tikz{\draw[black,solid,line width=0.9pt](0,0) -- (5mm,0);}}}
\newcommand{\REDsolid}{\raisebox{2pt}{\tikz{\draw[red,solid,line width=0.9pt](0,0) -- (5mm,0);}}}
\newcommand{\BLUsolid}{\raisebox{2pt}{\tikz{\draw[blue,solid,line width=0.9pt](0,0) -- (5mm,0);}}}
\newcommand{\GREsolid}{\raisebox{2pt}{\tikz{\draw[black!30!green,solid,line width=0.9pt](0,0) -- (5mm,0);}}}
\newcommand{\ORAsolid}{\raisebox{2pt}{\tikz{\draw[orange,solid,line width=0.9pt](0,0) -- (5mm,0);}}}

\newcommand{\REDdash}{\raisebox{2pt}{\tikz{\draw[red,dashed,line width=0.9pt](0,0) -- (5mm,0);}}}
\newcommand{\BLUdash}{\raisebox{2pt}{\tikz{\draw[blue,dashed,line width=0.9pt](0,0) -- (5mm,0);}}}
\newcommand{\GREdash}{\raisebox{2pt}{\tikz{\draw[black!30!green,dashed,line width=0.9pt](0,0) -- (5mm,0);}}}
\newcommand{\ORAdash}{\raisebox{2pt}{\tikz{\draw[orange,dashed,line width=0.9pt](0,0) -- (5mm,0);}}}

\newcommand{\REDdashdot}{\raisebox{2pt}{\tikz{\draw[red,dash dot,line width=0.9pt](0,0) -- (5mm,0);}}}
\newcommand{\GREdashdot}{\raisebox{2pt}{\tikz{\draw[black!30!green,dash dot,line width=0.9pt](0,0) -- (5mm,0);}}}

\newcommand{\REDdot}{\raisebox{2pt}{\tikz{\draw[red,dotted,line width=0.9pt](0,0) -- (5mm,0);}}}
\newcommand{\BLUdot}{\raisebox{2pt}{\tikz{\draw[blue,dotted,line width=0.9pt](0,0) -- (5mm,0);}}}
\newcommand{\GREdot}{\raisebox{2pt}{\tikz{\draw[black!30!green,dotted,line width=0.9pt](0,0) -- (5mm,0);}}}
\newcommand{\ORAdot}{\raisebox{2pt}{\tikz{\draw[orange,dotted,line width=0.9pt](0,0) -- (5mm,0);}}}
\title{Blinded sample size re-calculation in multiple composite population designs with normal data and baseline adjustments}

\DOIsuffix{}
\Volume{}
\Issue{}
\Year{}
\pagespan{1}{}

\title[Blinded sample size re-calculation in multiple composite populations]{Blinded sample size re-calculation in multiple composite population designs with normal data and baseline adjustments}
\author[Gera]{Roland G. Gera\footnote{Corresponding author: {\sf{e-mail: roland.gera@med.uni-goettingen.de}}}\inst{1}} 
\address[\inst{1}]{Department of Medical Statistics, University Medical Centre Göttingen, Göttingen, Germany}
\address[\inst{2}]{DZHK (German Center for CardiovascularResearch), Partner Site Göttingen, Göttingen, Germany}
\author[Friede]{Tim Friede\inst{1,2}}

\Receiveddate{ } \Reviseddate{ } \Accepteddate{ } 

\begin{abstract}
The increasing interest in subpopulation analysis has led to the development of various new trial designs and analysis methods in the fields of personalized medicine and targeted therapies. In this paper, subpopulations are defined in terms of an accumulation of disjoint population subsets and will therefore be called composite populations. The proposed trial design is applicable to any set of composite populations, considering normally distributed endpoints and random baseline covariates. Treatment effects for composite populations are tested by combining $p$-values, calculated on the subset levels, using the inverse normal combination function to generate test statistics for those composite populations. The family-wise type I error rate for simultaneous testing is controlled in the strong sense by the application of the closed testing procedure. Critical values for intersection hypothesis tests are derived using multivariate normal distributions, reflecting the joint distribution of composite population test statistics under the null hypothesis. For sample size calculation and sample size re-calculation multivariate normal distributions are derived which describe the joint distribution of composite population test statistics under an assumed alternative hypothesis. Simulations demonstrate the strong control of the family-wise type I error rate in fixed designs and re-calculation designs with blinded sample size re-calculation. The target power after sample size re-calculation is typically met or close to being met.
\end{abstract}

\maketitle                   






\section{Introduction}

In recent years, there has been an increased interest in evaluating the treatment effect across heterogeneous subpopulations \citep{Kent2010, Varadhan2013, Basu2017}. This approach has been coined the terms \emph{precision} or \emph{personalized} medicine. One issue is that the pooling of treatment effects coming from diverse subpopulations will lead to a blending of those treatment effects. When such a blended treatment effect is evaluated it might occur that a beneficial treatment effect for one subpopulation is diluted by subpopulations where the treatment benefit is not as high. This emphasizes the need for trial designs that aim to test treatment efficacy not only in the full population but in subpopulation as well. \\
One view of a subpopulation is to define it as an amalgamation of several population subsets. Those subsets are usually identified using biomarkers. Common examples for biomarkers include genetic markers but also patient characteristics such as age and gender. In relation to the outcome, a biomarker can be prognostic and therefore describe natural differences in the overall outcome between patients in different subsets and/or predictive, where a biomarker indicates that a treatment has different efficacy in different subsets \citep{Jenkins2011}. If at the planing stage for a trial heterogeneous reactions in regards to the treatment is expected, be it by subpopulation specific efficacies, safety profiles or any other clinical relevant behaviour, it makes sense to divide the study population into subpopulations where similar responses towards the treatment are expected. These populations should be pre-defined prior to the study \citep{Tanniou2016, Tanniou2019, MedicinesAgency2019}. \\
Incorporating heterogeneous study populations in a trial is not trivial and numerous challenges have to be addressed. One of these challenges is the interpretation of study results. The \textit{Points to consider for multiplicity issues of clinical trials} by the \citet{EMA1} for example points out that, from a regulatory perspective, a positive result in the overall population may not lead to valid claims in all subpopulations as long as there is reason to expect heterogeneity in the respective populations. Besides the issue of heterogeneity in study populations, another point to look out for when planning a study is how relevant covariates can be included in the study. Both, the \textit{Guideline on adjustment for baseline covariates in clinical trials} \citep{Points} as well as the ICH E9 support the inclusion of a priori defined baseline covariates in the analysis in cases where strong or moderate association between the covariates and the outcome is assumed. \\
Lastly, sample size calculation is an integral part of any trial planning phase. The sample size should be large enough to detect a certain relevant effect size while avoiding unnecessary large sample sizes out of ethical, time and monetary restrictions. However, the sample size is dependent on many parameters which are needed to define the final sample size. Many of those parameters are often not of actual research interest and are therefore called nuisance parameters \citep{NUISANCE1}. Estimates for any parameter is often based on either previous studies or by making educated guesses about the magnitude of the true parameter. When considering a trial with multiple composite populations and population subsets where the effect of a new treatment has to be determined, common nuisance parameters are the covariances between the endpoint and each covariate as well as the variance of the endpoint itself. These parameters have to be defined for each subset. A misspecification can lead to an inadequately sized study, either being too small and therefore leading to a study with lacking power or being to large and wasting the time and money of researcher and subjects alike. A sample size review during an internal pilot study can provide a solution to this problem \citep{Wittes1990}. The basic principle of a sample size review is to look into the data at an earlier stage of the trial, to re-estimate the nuisance parameters based on the available data and to re-calculate the required sample size. Generally, regulatory agencies prefer a blinded (or non-comparative) sample size review, i.e. without unveiling of the treatment allocation, over unblinded procedures \citep{EMA1998, EMEA2007}. When treatment allocations remain concealed the number of potential sources of bias are reduced. The editorial by \citet{Julious2015} gives a broad overview of works in regards to pilot studies. \\
Already, methods accounting for some, but not all, highlighted challenges have been presented. An approach by \citet{Mehta2014} describes a trial design for survival endpoints in cancer trials. There, only disjoint subsets were tested instead of overlapping subpopulations. Since test statistics are derived from disjoint subsets those test statistics are independent from each other which makes controlling the family-wise type I error rate (FWER) a simple matter. While this approach is quite accessible, it excludes any situations where overlapping populations exist. Using our method, trial designs with disjoint subsets as well as composite populations can be considered. \\
\citet{Placzek2017} presented a trial design to evaluate treatment effects in nested subpopulations while also including a sample size calculation and blinded sample size re-calculation scheme, considering continuous normally distributed endpoints. However, the use of covariates was not investigated and no trial design for non-nested subpopulations was provided. \\
\citet{Graf2018} investigated the issue where an continuous biomarker is dichotomized to identify subpopulations. If several biomarker thresholds have to be tested, the repeated testing of increasing thresholds raises the issue of multiple testing while at the same time resembling the structure of a group sequential trial. In the paper, various procedures which account for multiple testing ware investigated and their operating characteristics compared. \\
In contrast, the method by \citet{Chiu2018} shows an approach for testing multiple subpopulations at once. Additionally, this approach is then further extended to test those subpopulation structure within and multi-stage trial design framework. First, disjoint subsets are considered which are then combined into subpopulations using the inverse normal combination function. However, only normally distributed endpoints with equal and known variances in each subset are considered while also missing the use of covariates in the analysis. The testing approach presented in this paper is similar to the one by Chiu et al. but allows for varying variances and correlations in population subsets and includes covariates in the analysis while additionally showing how and effective sample size re-calculation can be conducted.\\
On the topic of sample size calculation, \citet{Friede2011} presented a method to re-determine the nuisance parameter during an internal pilot study and use this to conduct a blinded sample size re-calculation for ANCOVA designs with one normally distributed covariate. But while this approach was lately extended by \citet{Zimmermann2020} to account for multiple normally distributed covariates, it is not applicable for the subpopulation design. Here, we extend these ideas to blinded sample size re-calculation within multiple composite population settings.\\

\noindent This paper is structured as follows. In Section 2, we motivate our design by an example on pulmonary arterial hypertension. Section 3 describes our approach for constructing the statistical model and hypothesis testing using any composite population structure, while also presenting the sample size calculation and re-calculation schemes. Simulation results are presented in Section 4. The paper closes with our findings, a brief discussion and conclusions in Section 5.

\section{Motivating example}
The chronic and progressive disorder pulmonary hypertension (PH) is a rare disease which often occurs with non-specific symptoms like shortness of breath, fatigue, swelling of the legs, swelling of the ankles, chest pain and light-headedness. These symptoms are caused by an increased blood pressure in the pulmonary circulation system. Over time, this increased strain on the right heart side leads to a failure of that heart side which will ultimately results in death if the illness remains untreated \citep{Montani2013}. While the early symptoms seem mild and often only occur during physically demanding exercise, it is important to treat PH as early as possible. New treatments for this illness are being developed frequently, requiring novel and efficient trial designs to accompany these trials. \citet{Grieve2014} summarize results of a workshop where novel designs and other design options for PH trials were discussed. Some of the highlighted methods encompass blinded sample size re-calculation, subpopulation analysis and population enrichment designs. Especially the last two options lent themselves very well to trials in PH, since currently five main subtypes of PH are recognized which may be further divided. \citet{Simonneau2019} describe these categories, which were defined in more detail during the latest world symposium on PH. Pulmonary arterial hypertension (PAH) for example is divided according to idiopathic, heritable, drug and toxins or illness related reasons. The category of PH due to left heart disease has subsets like, PH due to left ventricular systolic/diastolic dysfunction, valvular disease or due to congenital/acquired left heart inflow/outflow tract obstruction and congenital cardiomyopathies. The third category, PH due to lung diseases and/or hypoxia, is a collective term for various lung diseases, known to trigger PH, like chronic obstructive pulmonary disease or sleep-disordered breathing for example. Chronic thromboembolic pulmonary hypertension is a rare form of PH which is caused by undissolving blood clots in the lungs, resulting in scare tissue developing in the small blood vessels of the lung which in turn increases the blood pressure. Lastly, PH with unclear and/or multi-factorial mechanisms is a classification for a group of diseases largely grouped into haematologic disorders, systemic disorders, metabolic disorders or other disorders. 

Since the cause of PH can have various reasons, a treatment could affect subjects differently. \citet{McLaughlin2009} provide arguments why subpopulation analysis is important in the context of those categories and sub-categories. They point out that in randomised trials for PAH more than 50\% of the trial patients suffered from idiopathic PAH while the remaining subject suffered from other forms. They report that most treatments they investigated showed efficacy for patients suffering from idiopathic PAH while the efficiency of the same treatments decreased for patients suffering from non-idiopathic PAH. Therefore, by blending the heterogeneous treatment effects of all PAH sub-categories the interpretation of the results becomes difficult. For further information, see \citet{McLaughlin2009}\\
Beside the problem of heterogeneous treatment effects, McLaughlin et al. also highlight that many commonly used primary endpoints considered during trials for PH, such as the 6-minute walk (6MW), are influenced by other factors such as age and height. Overall, composite population testing allows for flexible testing strategies when multiple subgroups are considered and often random covariates can be identified which are correlated to the study outcomes and should be included in the study.

\section{Design}

\subsection{Statistical model}
We consider a two-arm randomised controlled trial design where an active treatment ($T$) is tested against a comparator treatment ($C$). The total sample size of subjects in the full study population is denoted as $N$ while the treatment allocation is denoted by $\kappa = N_T/N_C$, with $N_T$ and $N_C$ denoting the sample sizes for the active and control treatments. We further assume that the full population ($F$) consists of $J$ disjoint subsets $S_j$ with subset sample sizes given by $N_{Tj}$ and $N_{Cj}$ respectively and prevalences defined as $\tau_j=N_{Tj} / N_T=N_{Cj} / N_C$ for $j=1, ..., J$.\\
Based on subsets $S_j$, we can define composite populations allowing for arbitrary population structures within the full population. Let $R$ denote the number of composite populations defined as $G_{\mathcal{I}_r} = \bigcup\limits_{j\in \mathcal{I}_r} S_j$ for an index set $I_r$ and $r=1, ..., R$. Examples for possible composite populations based on four subsets can be found in Figure 1.

\begin{figure}[htbp]
\begin{center}
\includegraphics[width = 0.7\textwidth]{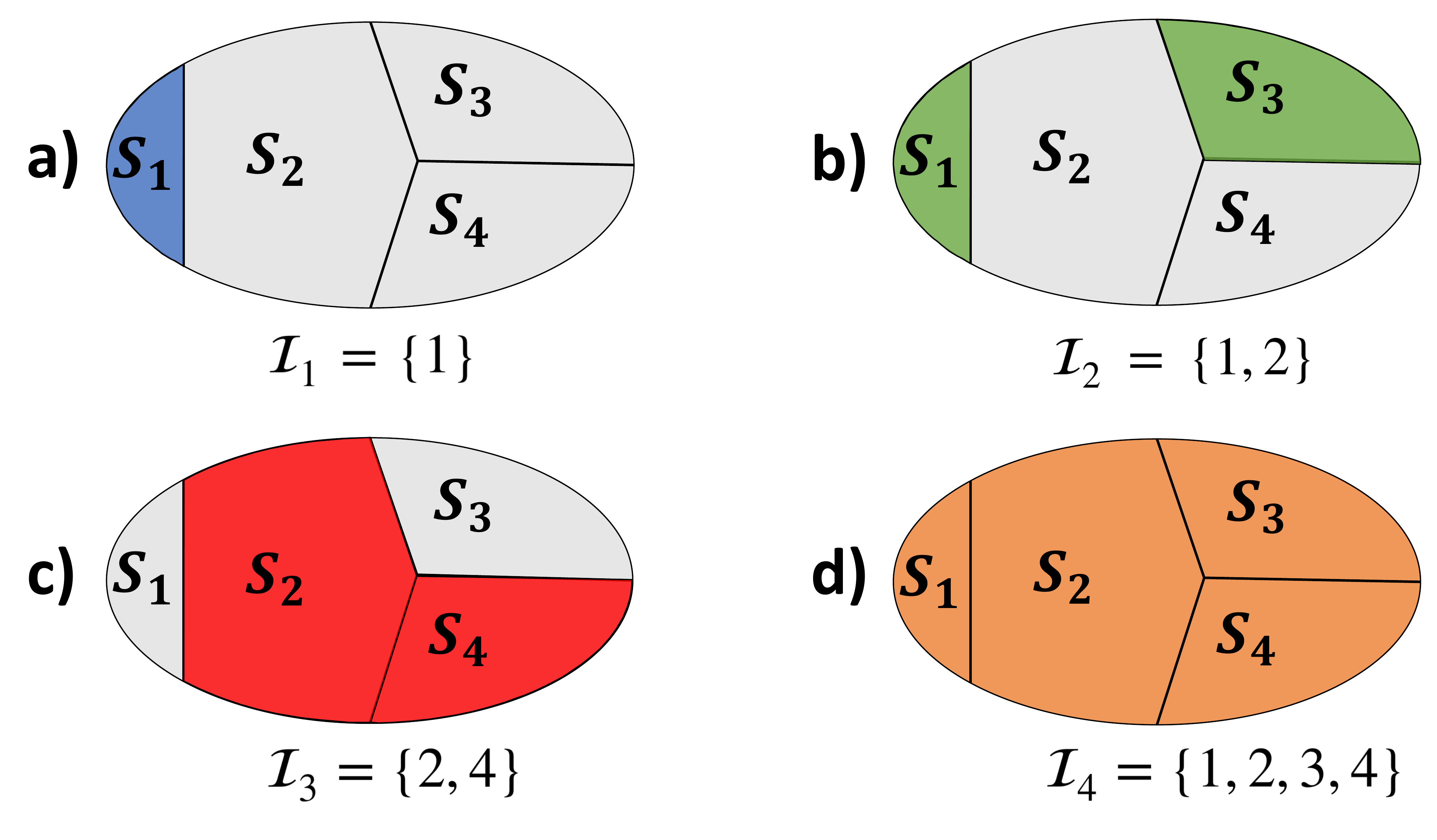}
\caption{Four possible composite population compilations for composite populations $G_{\mathcal{I}_1},\dotso , G_{\mathcal{I}_4}$, demonstrating how any composite population can be derived by combining available disjoint subsets.}
\label{fig:Pop}
\end{center}
\end{figure}

Then, for our study design we assume that endpoint $Y_j$ as well as $D$ random covariates $X_{jd}$, $d=1,\dotso,D$, were measured for each subject in each subset. For the covariates, no assumptions about their distributions are imposed, however, between each pair of variables a Pearson correlation coefficient is calculable. These correlations take the true values $\rho_{gqj}$, with $g,q \in \left\lbrace Y_j,X_{j1},\dotso,X_{jD} \right\rbrace$. Therefore, $D\cdot(D+1)/2$ distinct correlation coefficients are present in each subset. Due to randomisation we conclude that $\mu_{X_{Tjd}}=\mu_{X_{Cjd}}=\mu_{X_{jd}}$. Then, for any subject $k$ in subset $j$, we assume the following linear model 
\begin{align}
y_{kj} = \beta_{0j} +\beta_{1j} u_{ki} + \beta_{2j} x_{kj1}+\dotso+\beta_{(D+1)j}x_{kjD}+\epsilon_{kj},
\end{align}
where $k=1,\dotso,N_{Tj}+N_{Cj}$ and $j=1,\dotso,J$. The treatment indicator $u_{ki}$ equals 1 if $i=T$ and takes the value of 0 otherwise. All residuals are independent from each other and follow the distribution $\epsilon_{kj}\sim N(0,\sigma^2_{\epsilon_j})$ in any given subset $j$.\\
For simplicity, in this paper the one-sided subset null hypothesis $H_{0j}:\beta_{1j} \leq 0$ against its alternative $H_{Aj}:\beta_{1j} > 0$ is considered for testing. Therefore, it is assumed that greater values of $\beta_{1j}$ are associated with an increased treatment benefit. The null hypothesis $H_{0j}$ for treatment benefit $\beta_{1j}$ is tested using test statistic $T_j=\widehat{\beta}_{1j}/ S(\widehat{\beta}_{1j})$ where the variance for estimate $\widehat{\beta}_{1j}$ is determined by
\begin{align}
S^2 \left( \widehat{\beta}_{1j} \right) =& \frac{N_{Tj}+N_{Cj}-2}{N_{Tj}+N_{Cj}-2-D}\left(1-\widehat{\rho}_{Y_j,X_{j1}\dotso X_{jD}}^2\right)S^2_{Y_j}\cdot \nonumber \\
&\left( \frac{1}{N_{Tj}} + \frac{1}{N_{Cj}} + \mathscr{D}_{\overline{\mathbf{X}}_{Tj}-\overline{\mathbf{X}}_{Cj}}'\left(\left(N_{Tj}+N_{Cj}-2\right)\widehat{\Sigma}_{\mathbf{X}_j}\right)^{-1}\mathscr{D}_{\overline{\mathbf{X}}_{Tj}-\overline{\mathbf{X}}_{Cj}} \right).
\end{align} 
Let $\widehat{\rho}_{Y_j,X_{j1}\dotso X_{jD}}$ denote the coefficient of multiple correlation. Similar to how a Pearson correlation coefficient measures the linear association between one random variable and another, the coefficient of multiple correlation measures the linear association between an ensemble of random variables and one other variable \citep{Cohen2003}. Here, it measures the linear association between the variables $X_{j1}\dotso X_{jD}$ and the endpoint $Y_j$. For the $1\times D$ column vector $\mathscr{D}_{\overline{\mathbf{X}}_{Tj}-\overline{\mathbf{X}}_{Cj}}$, each entry of the vector represents the calculated difference between means $\overline{X}_{djT}$ and $\overline{X}_{djC}$, therefore $\overline{X}_{djT}-\overline{X}_{djC}$. Finally, $\widehat{\Sigma}_{\mathbf{X}_j}$ denotes a $D\times D$ covariance matrix calculated for the covariates $X_{j1},\dotso,X_{jD}$.\\
Given $H_{0j}$, subset test statistic $T_j$ follows a central t-distribution with $N_{Tj}+N_{Cj}-2-D$ degrees of freedom. A $p$-value testing the one-sided $H_{0j}$ is calculated by $p_j=1-\Psi_{N_{Tj}+N_{Cj}-2-D}\left( T_j \right)$, where $\Psi_{\text{df}}$ denote the cumulative distribution function for a t-distribution with df degrees of freedom with the sole restriction that for each subset df $\geq 1$ must be true. \\
To derive test statistics for composite populations, $p$-value combination functions are applied. To merge multiple independent subset $p$-values, we use the inverse normal combination function \citep{Lehmacher1999}. Hence, a test statistic is needed to test the composite population null hypothesis $H_{0G_{\mathcal{I}_r} }$ against its alternatives. \citet{Birnbaum1954} defines the null hypothesis for such composite populations as 
\begin{align*}
&H_{0G_{\mathcal{I}_r} }:\text{all }\beta_{1j}\leq 0, j\in \mathcal{I}_r\\
&H_{AG_{\mathcal{I}_r} }:\text{one or more } \beta_{1j}>0, j\in \mathcal{I}_r.
\end{align*} 
Hence, the test statistics for the composite null hypotheses above are calculated as
\begin{align}
Z_{G_{\mathcal{I}_r}} =  \sum_{j \in \mathcal{I}_r} \sqrt{\frac{w_j}{\sum\limits_{k \in \mathcal{I}_r} w_k}}  \Phi^{-1}(1-p_j).
\end{align}
Here, $\Phi^{-1}(\cdot)$ denotes the quantile function of the standard normal distribution. Let $w_j$ define pre-fixed weights for subset $S_j$ with $w_j>0$ for all $j=1,\dotso,J$ and $\sum_{j=1}^J w_j^2=1$. Those weights are often a function of sample size or pre-estimated standard errors. Since $p$-values are derived for each subset separately and combined afterwards, parameters like variances and correlations can vary between subsets without compromising the FWER and without diluting these parameters.\\
Invoking the \textit{p-clud} property as described by \citet{Brannath2009} it can be shown that if $H_{0G_{\mathcal{I}_r} }$ is true and the \textit{p-clud} property is satisfied, then $1-\Phi(Z_{G_{\mathcal{I}_r}})$ does not exceed any given $\alpha$ level. The \textit{p-clud} property is fulfilled if the distribution of the $p$-value $p_j$, $j\in \mathcal{I}_r$, and the conditional distributions of $p_{j'}$ given $p_j$, $j'\in \mathcal{I}_r\setminus j$, are stochastically larger then or equal to the uniform distribution on [0,1]. Here, using the t-test statistic, this condition is always satisfied, since the subset $p$-values emerge from disjoint subsets and are therefore always independent from each other.\\ 
Since $R$ composite populations are being tested, multiple testing needs to be accounted for. We aim to control the FWER in a strong sense by using the closed testing principle \citep{Marcus1976,Brannath2009}, meaning that any given nominal level $\alpha$ acts as an upper boundary for the FWER. To this end, intersection hypotheses are defined as $H_{0\cap_{r\in \mathcal{K}}G_{\mathcal{I}_r} } = \cap_{r\in \mathcal{K}}H_{0G_{\mathcal{I}_r} }$, where $\mathcal{K}\subseteq \left\lbrace 1,\dotso,R \right\rbrace$. Strong control is achieved, if null hypothesis $H_{0G_{\mathcal{I}_r} }$ is only rejected at nominal level $\alpha$ provided that all intersection hypotheses containing the null hypothesis $H_{0G_{\mathcal{I}_r} }$ have been rejected at level $\alpha$ as well. To test an intersection null hypothesis approaches like the Bonferroni correction or the \v Sid\'ak correction could be applied \citep{Bretz2006, Schmidli2006, Friede2012, Friede2020}. These corrections, however, do not account for the correlation between composite population test statistics $Z_{G_{\mathcal{I}_r}}$. Correlations between composite test statistics occur if the same subsets are part of two or more composite populations. An approach that accounts for these correlations between completely nested, partially nested or disjoint composite populations was presented by \citet{Spiessens2010}. There, a multivariate normal distribution is used to describe the joint distribution of the test statistics given a true intersection null hypothesis. \\
Using this approach, distributions for the test statistics under the null hypothesis and corresponding critical values can be defined which fulfil any given $\alpha$ level. By accounting for the covariances between test statistics, more efficient testing strategies emerge. For instance, given intersection hypothesis $H_{0\cap_{r=1}^RG_{\mathcal{I}_r} }$, the vector of composite population test statistics $\mathbf{Z}$, $\mathbf{Z}=\left( Z_{G_{\mathcal{I}_1}},\dotso,Z_{G_{\mathcal{I}_R}} \right)$ follows the distribution
\noindent
\begin{align}
\mathbf{Z} \sim \mathcal{N}_R\left(  \mathbf{0} , \Sigma_0\right) 
\end{align}
with 
\begin{align*}
\Sigma_0 =  \left( \begin{array}{cccc}
1 & \text{cov}\left(Z_{G_{\mathcal{I}_1} },Z_{G_{\mathcal{I}_2} }\right) & \cdots & \text{cov}\left(Z_{G_{\mathcal{I}_1} },Z_{G_{\mathcal{I}_{R}} }\right) \\ 
\text{cov}\left(Z_{G_{\mathcal{I}_1} },Z_{G_{\mathcal{I}_2} }\right) & 1 & \ddots & \vdots \\ 
\vdots & \ddots & 1 & \text{cov}\left(Z_{G_{\mathcal{I}_{R-1}} },Z_{G_{\mathcal{I}_R} }\right) \\ 
\text{cov}\left(Z_{G_{\mathcal{I}_1} },Z_{G_{\mathcal{I}_R} }\right) & \cdots & \text{cov}\left(Z_{G_{\mathcal{I}_{R-1}} },Z_{G_{\mathcal{I}_R}} \right) & 1
\end{array}  \right) 
\end{align*}
Matrix $\Sigma_0$ denotes the covariance matrix between composite population test statistics given that the null hypothesis is true. The covariance between any two composite populations is solely dependent on the overlapping subsets that are present in both. To show the calculation of the covariances, let $Z_{G_{\mathcal{I}_r}}$ and $Z_{G_{\mathcal{I}_{r'}}}$ be two composite population test statistics with $r \neq r'$. Then the covariance between these two test statistics is calculated as
\begin{align*}
\text{cov}\left( Z_{G_{\mathcal{I}_{r}}},Z_{G_{\mathcal{I}_{r'}}} \right) &= \text{cov}\left(  \sum_{j \in \mathcal{I}_{r}} \sqrt{\frac{w_j}{\sum\limits_{k \in \mathcal{I}_{r}} w_k}}  \Phi^{-1}\left(1-p_j\right),  \sum_{j' \in \mathcal{I}_{r'}} \sqrt{\frac{w_{j'}}{\sum\limits_{k' \in \mathcal{I}_{r'}} w_{k'}}}  \Phi^{-1}\left(1-p_{j'}\right) \right).
\end{align*} 
This expression can be simplified however since composite population test statistics can be divided into the part where the same subsets exist in both composite populations and the part where subsets only exist in only one of the composite populations. This leads to 
\begin{align}
\text{cov}\left( Z_{G_{\mathcal{I}_{r}}},Z_{G_{\mathcal{I}_{r'}}} \right) =&\text{ cov}\left(  \sum_{j \in \mathcal{I}_r\cap\mathcal{I}_{r'}} \sqrt{\frac{w_j}{\sum\limits_{k \in \mathcal{I}_{r}} w_k}}  \Phi^{-1}\left(1-p_j\right),\sum_{j' \in \mathcal{I}_r\cap\mathcal{I}_{r'}} \sqrt{\frac{w_{j'}}{\sum\limits_{k' \in \mathcal{I}_{r'}} w_{k'}}}  \Phi^{-1}\left(1-p_{j'}\right)\right)+\nonumber\\
&\text{ cov}\left(  \sum_{j \in \mathcal{I}_r\cap\mathcal{I}_{r'}} \sqrt{\frac{w_j}{\sum\limits_{k \in \mathcal{I}_{r}} w_k}}  \Phi^{-1}\left(1-p_j\right), \sum_{h' \in \mathcal{I}_{r'}\setminus \mathcal{I}_{r}} \sqrt{\frac{w_{h'}}{\sum\limits_{k' \in \mathcal{I}_{r'}} w_{k'}}}  \Phi^{-1}\left(1-p_{h'}\right)\right)+ \nonumber\\
&\text{ cov}\left(  \sum_{h \in \mathcal{I}_r\setminus \mathcal{I}_{r'}} \sqrt{\frac{w_h}{\sum\limits_{k \in \mathcal{I}_{r}} w_k}}  \Phi^{-1}\left(1-p_h\right),\sum_{j' \in \mathcal{I}_r\cap\mathcal{I}_{r'}} \sqrt{\frac{w_{j'}}{\sum\limits_{k' \in \mathcal{I}_{r'}} w_{k'}}}  \Phi^{-1}\left(1-p_{j'}\right)\right)+\nonumber\\
&\text{ cov}\left(  \sum_{h \in \mathcal{I}_r\setminus \mathcal{I}_{r'}} \sqrt{\frac{w_h}{\sum\limits_{k \in \mathcal{I}_{r}} w_k}}  \Phi^{-1}\left(1-p_h\right), \sum_{h' \in \mathcal{I}_{r'}\setminus \mathcal{I}_{r}} \sqrt{\frac{w_{h'}}{\sum\limits_{k' \in \mathcal{I}_{r'}} w_{k'}}}  \Phi^{-1}\left(1-p_{h'}\right)\right).
\end{align} 
Removing the terms corresponding to covariances of disjoint subsets, (5) reduces to
\begin{align}		
\text{cov}\left( Z_{G_{\mathcal{I}_{r}}},Z_{G_{\mathcal{I}_{r'}}} \right)&= \sum_{l \in \mathcal{I}_r\cap\mathcal{I}_{r'}} \text{cov}\left( \sqrt{\frac{w_l}{\sum\limits_{k \in \mathcal{I}_{r}} w_k}}  \Phi^{-1}(1-p_l),  \sqrt{\frac{w_l}{\sum\limits_{k' \in \mathcal{I}_{r'}} w_{k'}}}  \Phi^{-1}(1-p_l) \right)\nonumber\\
			&= \sum\limits_{l \in \mathcal{I}_r\cap\mathcal{I}_{r'}} \sqrt{\frac{w_l}{\sum\limits_{k \in \mathcal{I}_{r}} w_k}} \sqrt{\frac{w_l}{\sum\limits_{k' \in \mathcal{I}_{r'}} w_{k'}}}  \text{cov}\left(  \Phi^{-1}(1-p_l),  \Phi^{-1}(1-p_l) \right)\nonumber\\
			&= \sum\limits_{l \in \mathcal{I}_r\cap\mathcal{I}_{r'}} \frac{w_l}{\sqrt{\sum\limits_{k \in \mathcal{I}_{r}} w_k \sum\limits_{k' \in \mathcal{I}_{r'}} w_{k'}}} \text{var}\left( \Phi^{-1}(1-p_l) \right)\nonumber\\
			&= \sum\limits_{l \in \mathcal{I}_r\cap\mathcal{I}_{r'}} \frac{w_l}{\sqrt{\sum\limits_{k \in \mathcal{I}_{r}} w_k \sum\limits_{k' \in \mathcal{I}_{r'}} w_{k'}}} .
\end{align} 
The covariance between any two $Z_{G_{\mathcal{I}_{r}}}$ and $Z_{G_{\mathcal{I}_{r'}}}$ can therefore be calculated using (6). This in turn is then applied to calculate each entry in $\Sigma_0$. This gives the covariance structure for the distribution of the intersection hypothesis. By using the joint distribution in (4), $R$ critical values can be calculated so that the cumulative distribution function, given those critical values lead to an error rate of $\alpha$ so that equation
\begin{align}
\alpha \geq P_{H_{0\cap_{r=1}^RG_{\mathcal{I}_r} }}\left( Z_{G_{\mathcal{I}_{1}}}\geq c_1 \vee \dotso \vee Z_{G_{\mathcal{I}_{R}}}\geq c_R \right)
\end{align}
is fulfilled. There exists an infinite number of sets containing $R$ critical values which fulfil (7). However, by requiring that the same critical values, $c_G$ is used, only one value for $c_G$ exists which fulfils equation
\begin{align}
\alpha \geq P_{H_{0\cap_{r=1}^RG_{\mathcal{I}_r} }}\left( Z_{G_{\mathcal{I}_{1}}}\geq c_G \vee \dotso \vee Z_{G_{\mathcal{I}_{R}}}\geq c_G \right).
\end{align}
Therefore, the $\alpha$-level and the burden of rejecting the null hypothesis is distributed equally among composite populations. After $c_G$ has been determined it is then applied to the testing of the intersection hypothesis by comparing each composite population test statistic to $c_G$. If even a single composite population is larger than $c_G$ the intersection null hypothesis is rejected and testing is continued on a lower testing level.

\subsection{Sample size calculation}
For the sample size calculation several different power definitions could be used. Here, we opted to apply the so called \textit{disjunctive power}. The disjunctive power is to the probability to reject at least one false null hypothesis \citep{Senn2007}. The required sample size for a study depends on subset prevalences, variances and correlations. However, those parameters are typically not of research interest. For an initial sample size calculation, assumptions on the effect sizes $\beta_{1j}$, variances $\sigma^2_{Y_j},\sigma^2_{X_{j1}},\dotso, \sigma^2_{X_{jD}}$ and the correlations $\rho_{gqj}$ have to be made and while often a lower limit of effectiveness can be defined for the treatment benefit $\beta_{1j}$, typically no such reference points can be defined for the other parameters. Hence, previous studies have to be consulted to get estimates for those nuisance parameters. If no such studies exist, values for the nuisance parameters have to be estimated without previous knowledge. In the following, test statistics based on prior assumptions and an arbitrary but fixed overall sample size $N$ are denoted as $Z_{G_{\mathcal{I}_r}}^*$. Using those assumptions a joint test statistic under the alternative hypothesis is derived similar to the joint test statistic for the null hypothesis in (4). Since a treatment benefit is assumed to exist in at least one subset, the distribution of the joint test statistic consists of non-zero mean vector $\left( Z_{G_{\mathcal{I}_1}}^*,\dotso,Z_{G_{\mathcal{I}_R}}^* \right)$ and a covariance structure under the alternative hypothesis $\Sigma_A^*$. Let the joint cumulative distribution under the alternative be specified by $\mathbf{G}^* = \mathcal{N}_R\left( \left( Z_{G_{\mathcal{I}_1}}^*,\dotso,Z_{G_{\mathcal{I}_R}}^* \right),\Sigma_A^* \right)$. Naturally, the covariance structure under the alternative hypothesis, $\Sigma_A^*$, is different from the covariance structure under null hypothesis $\Sigma_0$. This difference increases with increasing treatment effect. Since no closed formula to determine $\Sigma_A^*$ is available, we approximate $\Sigma_A^*$ by simulating patient data using our initial assumptions. Then, subset $p$-values and composite population test statistics are calculated following the planned study design. By doing this procedure repeatedly, $R$ distinct lists consisting of composite test statistics are generated. From these lists, the covariance between any two test statistics can be calculated which determines $\widehat{\Sigma}_A$. Since the values in $\Sigma_A^*$ are unaffected by sample sizes, the determination of $\Sigma_A^*$ has to be done only once, namely at the beginning of the sample size calculation process. The more entries the lists of the composite test statistics have and the more simulated subjects each simulation run has, the more precise the estimation of $\widehat{\Sigma}_A$ becomes. In the next step, the estimated $\widehat{\Sigma}_A$ replaces $\Sigma_A^*$ in the joint cumulative distribution $\mathbf{G}^*$. With larger sample sizes the vector of mean differences $\mathscr{D}_{\overline{\mathbf{X}}_{Tj}-\overline{\mathbf{X}}_{Cj}}$ converges to a zero vector and is therefore ignored in the following sample size considerations. Hence from here on out, formula (2) is simplified to $S^2(\widehat{\beta}_{1j})\approx(N_{Tj}+N_{Cj}-2)/(N_{Tj}+N_{Cj}-2-D)\cdot\left(1-\widehat{\rho}_{Y_j,X_{j1}\dotso X_{jD}}^2\right)S^2_{Y_j}\cdot\left( \frac{1}{N_{Tj}} + \frac{1}{N_{Cj}}\right)$.\\
Then, the required sample size is calculated by determining the predicted joint composite population test statistic given the initially assumed values. Since all other variables remain fixed we have to find a sample size $N$ which fulfils
\begin{align}
1-\beta \geq P_{H_{A\cap_{r=1}^RG_{\mathcal{I}_r} }}\left( Z_{G_{\mathcal{I}_{1}}}^*\geq c_G\vee \dotso \vee Z^*_{G_{\mathcal{I}_{R}}}\geq c_G | N\right).
\end{align}
The smallest $N$ which satisfies (9) is denoted by $N_0$. Equally, (9) can be written as
 \begin{align*}
N_0 = \text{min } N \quad \text{ s.t.  } \quad 1 - \mathbf{G}^*(c_G) \geq 1-\beta.
\end{align*}
However, lower bounds for $N_0$ do apply, namely that $N_0\cdot \tau_j -2-D\geq 1$ has to be true for all subsets $j$. The calculation for critical values can simply be done using \texttt{R} software packages like \texttt{mvtnorm} \citep{MVTNORM,MVTNORM1} or by utilizing the cumulative density function for multivariate normal distributions implemented in \texttt{SAS} or \texttt{MATLAB}. \\

\subsection{Sample size re-calculation using the internal pilot study design}
During the initial planning stage researchers are basing their justification for the sample size on prior assumptions of the nuisance parameters which could be wrong, inaccurate or not be applicable in the current study population. Therefore, it seems sensible to stop the trial at a certain point during the recruitment phase and to re-estimate nuisance parameters based on collected data. Such a procedure is executed during an internal pilot study (IPS) \citep{Wittes1990} and can be broken down into three steps. The first step is to calculate the initial global sample size $N_0$ based on initial assumptions about the nuisance parameters. For the second step, let $\nu$ denote a fraction of the initially planned sample size. After $\nu\cdot N_0$ subjects are recruited, an internal pilot study is conducted wherein the nuisance parameters are re-estimated based on the newly collected data. The number of patients at which the pilot study takes place is defined as $N_1$, where $N_1 = \nu\cdot N_0$. An overview of how to size an internal pilot study can be found in \citet{Friede2006} as well as in \citet{Friede20102}. Based on the $N_1$ subjects parameters are re-estimated without unveiling treatment affiliations and those parameters are then used to conduct a blinded sample size re-calculation (BSSR). To estimate the nuisance parameters of the linear models without unveiling the treatment allocation of the subjects, we are building upon the framework of \citet{Zimmermann2020}. They describe methods to blindly re-estimate the variance for the adjusted treatment effects when testing a normally distributed outcome with an arbitrary number of normally distributed random baseline covariates. That approach is an extension of the work by \citet{Friede2011}. \\
Hence, after $N_1$ subjects have been recruited the sample size review is conducted using the same linear regression model as in (1) This way, the treatment allocations of all subjects remain concealed. The model is evaluated in each subset, using all available data. However, there exists a minimum sample size for $N_1$ where the equation $N_1\cdot \tau_j -1-D\geq 1$ has to be fulfilled for every single subset $j$. Based on data from $N_1$ subjects $J$ linear regression models are fitted for the blinded re-calculation. The models are similar to (1) with the only difference being that $\beta_{1j}u_{ki}$ is omitted. Therefore, the treatment allocation remains concealed. The estimated residuals $\widehat{\epsilon}_{kj}$ of these models are then utilized to approximate the variances for $\widehat{\beta}_{1j}$ by
\begin{align*}
\frac{N_{Tj}+N_{Cj}-2}{N_{Tj}+N_{Cj}-2-D}\left(1-\widehat{\rho}_{Y_j,X_{j1}\dotso X_{jD}}^2\right)S^2_{Y_j}\approx \frac{\sum_{k=1}^{N_{Tj}+N_{Cj}}\widehat{\epsilon}_{kj}^2}{N_{Tj}+N_{Cj}-1-D}.
\end{align*}
This approximation is then applied to formula (2) and the sample size calculation procedure is repeated with the updated estimations resulting in a re-calculated sample size $N_{\text{reest}}$. However, since the treatment allocation is ignored, a bias is introduced into the variance estimators leading to slightly inflated estimators. This is often not a problem since it has been shown for ANCOVA models with one random covariate, that the impact of this inflation is generally negligible (see \citet{Friede2013}). \\
The third and last step is to decide which re-calculated sample size to use as the final sample size$N_{\text{final}}$. Wittes and Britain proposed the so called restricted design where the final sample size is defined as the maximum of the initially calculated sample size and the re-calculated sample size, therefore $N_{\text{final}}=\text{max}(N_0,N_{\text{reest}})$. \citet{Birkett1994} on the other hand proposed the unrestricted design, where the sample size is the maximum of the sample size at the internal pilot study and the re-calculated sample size, hence $N_{\text{final}}=\text{max}(N_1,N_{\text{reest}})$. In this design, should the re-calculated sample size be lower than $N_1$ then the final analysis will be based on $N_1$ patients.

\section{Simulation study}

The performance of the proposed procedures were evaluated using simulation studies. For simplicity, a population consisting of a subsets $S_1$ and its complement $S_2$ was considered. The outcomes and covariates were drawn from a bivariate normal distribution with mean vector $(\beta_{11},0)'$ for treatment patients in subset $S_1$, while mean vector $(0,0)'$ was applied for the control subjects of $S_1$ and all subjects of $S_2$. The respective covariance matrix was defined by the corresponding scenario. The assumed prevalence $\tau_1^*$ always matched the true prevalence $\tau_1$. The effect of the intervention was tested in composite populations $G_{I_1}=S_1$ and $G_{I_2}=S_1\cup S_2$. Sample size calculation and re-calculation was performed such that it should match a disjunctive power of $1-\beta$ = 0.9. Results for scenarios corresponding to a disjunctive power of $1-\beta$ = 0.8 are found in the appendix. Further details on each scenario are shown in Table 1.

\begin{table}[h!]
\centering
\caption{Scenarios considered for the FWER and power simulation studies}
\label{tab:my-table}
\begin{tabular}{c|cc}
 & \shortstack{Under intersection\\null hypothesis $H_{0\cap_{r=1}^2G_{\mathcal{I}_r} }$ }& \shortstack{Under intersection \\alternative hypothesis $H_{A\cap_{r=1}^2G_{\mathcal{I}_r} }$} \\\toprule
 Simulation runs                         & 10 000                                 & 10 000 \\
Significance level $\alpha$                         & 0.025 (one-sided)                                 & 0.025 (one-sided)                                        \\
Allocation ratio $\kappa$                           & 1                                                 & 1                                                        \\
Prevalence $\tau_1^*=\tau_1$                      & 0.25; 0.5; 0.75                                          & 0.25; 0.5; 0.75 \\
Proportion $\nu$ for IPS                             & 0.3; 0.5                                          & 0.3; 0.5                                                 \\
Assumed treatment benefit $\beta_{11}^*$ in $S_1$  & 0.5; 1                                            & 0.5; 1                                                   \\
Assumed treatment benefit $\beta_{12}^*$  in $S_2$                        & 0                                                 & 0                                                        \\
Assumed variance $\sigma_{Y_1}^{2*}$ in $S_1$                       & 1                                                 & 1                                                        \\
Assumed variance $\sigma_{Y_2}^{2*}$ in $S_2$                    & 1                                                 & 1                                                        \\
   Assumed squared correlation $\rho_{Y_1,X_1}^{2*}=\rho_{Y_2,X_2}^{2*}$                                             & 0.4                                               & 0.4                                                      \\
True treatment benefit $\beta_{11}$ in $S_1$       & 0                                                 & 0.5; 1                                                   \\
True treatment benefit $\beta_{12}$   in $S_2$                                   & 0                                                 & 0                                                        \\
True variance $\sigma_{Y_1}^2$ in $S_1$                            & 0.8; 1; 1.2                                       & 0.8; 1; 1.2                                              \\
True variance $\sigma_{Y_2}^2$  in $S_2$                            & 1                                                 & 1                                                        \\
  True  squared correlation  $\rho_{Y_1,X_1}^2=\rho_{Y_2,X_2}^2$                                             & 0.0; 0.4; 0.8                           & 0.0; 0.2; 0.4; 0.6; 0.8                                 
\end{tabular}
\end{table}
\sbox3{\REDsolid}\sbox6{\REDdash}\sbox1{\BLUsolid}\sbox4{\BLUdash}\sbox5{\GREdash}\sbox2{\GREsolid}\sbox0{\BLAsolid}\sbox7{\GREdot}\sbox8{\BLUdot}
To evaluate the ability of the proposed blinded sample size re-calculation to maintain the desired power, we performed simulations of the power, the mean recalculated sample size and the variation of the re-calculated sample size. The variation was represented in terms of the 10\% as well as the 90\% quantile. The results are found in Figures 2 and 3. For the fixed designs (\usebox3) the parameters for the initial sample size calculation were assumed to be $\beta_{11}^*$, $\beta_{12}^*$, $\sigma_{Y_1}^{2*}$, $\sigma_{Y_2}^{2*}$ and $\rho_{Y_1,X_1}^{2*}=\rho_{Y_2,X_2}^{2*}$. The other two scenarios included a sample size re-calculation at a later stage with $\nu=0.5$ (\usebox1) or an earlier stage with $\nu=0.3$ (\usebox2). The expected standard error for these simulations is given by $\sqrt{\frac{0.1\cdot 0.9}{10000}}=0.003$. During the simulations, the true prevalences were used. \\

\begin{table}[htbp]
\caption{Comparing estimated probabilities of rejection of the intersection hypothesis $H_{0G_{\mathcal{I}_1}\cap G_{\mathcal{I}_2} }$ for fixed design and designs with blinded sample size re-calculation. The true and assumed values can be found in Table 1. Let $N_0$ denote the initially calculated total sample size using assumed variance and correlation but true prevalences. The predicted 95\%-confidence interval is calculated as [0.02194;0.02806] given a nominal level of $\alpha=0.025$. Values which fall out of the 95\%-confidence interval are highlighted. Out of the 162 simulations, 9 where out of bounds.} 
\centering
\begin{tabular}{ccccccccccc}
\toprule
  & &  \multicolumn{1}{l}{}& \multicolumn{4}{l}{$\mathbf{\beta_{11}=0.5}$} & \multicolumn{4}{l}{$\mathbf{\beta_{11}=1}$}\\
$\tau_1$ & $\sigma_{Y_1}^2$ &\multicolumn{1}{l}{ $\rho_{Y_1,X_1}^2=\rho_{Y_2,X_2}^2$} & $N_0$ &  \shortstack{FWER\\ no IPS} &  \shortstack{FWER\\ $\nu=0.3$} & \shortstack{FWER\\ $\nu=0.5$} &\multicolumn{1}{l}{$N_0$} &  \shortstack{FWER\\ no IPS} &  \shortstack{FWER\\ $\nu=0.3$} & \shortstack{FWER\\ $\nu=0.5$}  \\
\midrule
\rowcolor{gray!6}  0.25 & 0.8 & 0.0 & 648 & 0.0253 & 0.0266 & 0.0248 & 169 & 0.0237 & 0.0273 & 0.0262\\
\rowcolor{gray!6}       &     & 0.4 &     & 0.0240 & 0.0252 & 0.0259 &     & 0.0264 & 0.0246 & 0.0272\\
\rowcolor{gray!6}       &     & 0.8 &     & 0.0255 & 0.0236 & 0.0214 &     & 0.0243 & 0.0226 & 0.0263\\
\rowcolor{gray!6}       & 1   & 0.0 &     & 0.0262 & 0.0248 & 0.0240 &     & 0.0253 & 0.0267 & 0.0236\\
\rowcolor{gray!6}       &     & 0.4 &     & 0.0271 & 0.0256 & 0.0251 &     & 0.0245 & 0.0240 & 0.0269\\
\rowcolor{gray!6}       &     & 0.8 &     & 0.0262 & 0.0277 & 0.0233 &     & 0.0256 & \textbf{0.0281} & 0.0241\\
\rowcolor{gray!6}       & 1.2 & 0.0 &     & 0.0256 & 0.0272 & 0.0258 &     & 0.0257 & 0.0226 & 0.0247\\
\rowcolor{gray!6}       &     & 0.4 &     & 0.0249 & 0.0251 & 0.0232 &     & 0.0264 & 0.0225 & 0.0247\\
\rowcolor{gray!6}       &     & 0.8 &     & 0.0255 & 0.0267 & 0.0257 &     & 0.0285 & 0.0234 & \textbf{0.0289}\\
\addlinespace
0.50 & 0.8 & 0.0 & 313 & 0.0273 & 0.0245 & 0.0236 & 83 & 0.0250 & 0.0276 & 0.0240\\
     &     & 0.4 &     & 0.0266 & 0.0263 & 0.0238 &    & 0.0253 & 0.0257 & 0.0265\\
     &     & 0.8 &     & 0.0254 & 0.0242 & 0.0242 &    & \textbf{0.0208} & 0.0270 & 0.0239\\
     & 1   & 0.0 &     & 0.0250 & 0.0244 & 0.0235 &    & 0.0261 & 0.0225 & 0.0240\\
     &     & 0.4 &     & 0.0246 & 0.0266 & \textbf{0.0218} &    & 0.0258 & 0.0253 & 0.0251\\
     &     & 0.8 &     & 0.0270 & 0.0232 & 0.0239 &    & 0.0258 & 0.0222 & 0.0255\\
     & 1.2 & 0.0 &     & 0.0242 & 0.0252 & 0.0238 &    & 0.0264 & 0.0239 & 0.0260\\
     &     & 0.4 &     & 0.0261 & 0.0240 & 0.0240 &    & 0.0258 & 0.0236 & 0.0233\\
     &     & 0.8 &     & 0.0249 & 0.0266 & 0.0256 &    & 0.0243 & 0.0249 & 0.0228\\
\addlinespace
\rowcolor{gray!6}  0.75 & 0.8  & 0.0 & 201 & 0.0241 & 0.0252 & 0.0245 & 53 & 0.0237 & 0.0252 &\textbf{ 0.0219}\\
\rowcolor{gray!6}       &      & 0.4 &     & 0.0247 & \textbf{0.0215} & 0.0244 &    & 0.0258 & 0.0259 & 0.0272\\
\rowcolor{gray!6}       &      & 0.8 &     & 0.0226 & 0.0249 & 0.0239 &    & 0.0242 & 0.0237 & 0.0252\\
\rowcolor{gray!6}       & 1    & 0.0 &     & \textbf{0.0217} & \textbf{0.0213} & 0.0251 &    & 0.0280 & 0.0272 & 0.0227\\
\rowcolor{gray!6}       &      & 0.4 &     & 0.0241 & 0.0251 & 0.0232 &    & 0.0260 & 0.0244 & 0.0247\\
\rowcolor{gray!6}       &      & 0.8 &     & 0.0232 & 0.0241 & 0.0247 &    & 0.0222 & 0.0243 & 0.0231\\
\rowcolor{gray!6}       & 1.2  & 0.0 &     & 0.0254 & 0.0229 & 0.0247 &    & 0.0248 & 0.0252 & 0.0243\\
\rowcolor{gray!6}       &      & 0.4 &     & 0.0247 & 0.0227 & \textbf{0.0291} &    & 0.0252 & 0.0229 & 0.0265\\
\rowcolor{gray!6}       &      & 0.8 &     & 0.0229 & 0.0246 & 0.0263 &    & 0.0249 & 0.0260 & 0.0257\\
\bottomrule
\end{tabular}
\end{table}

\sbox3{\REDsolid}\sbox6{\REDdash}\sbox1{\BLUsolid}\sbox4{\BLUdash}\sbox5{\GREdash}\sbox2{\GREsolid}\sbox0{\BLAsolid}\sbox7{\GREdot}\sbox8{\BLUdot}
First the control of the FWER is investigates comparing the error rates of fixed designs and designs with earlier($\nu=0.3$) and later ($\nu=0.5$) sample size re-calculation. The sample sizes are calculated using assumed treatment benefits $\mathbf{\beta_{11}=0.5}$ and $\mathbf{\beta_{11}=1}$. Table 2 shows the results for the simulations given the FWER for low and large sample sizes considering a planed power of 90\%. The expected Monte-Carlo errors was calculated as $\sqrt{\frac{0.025\cdot 0.975}{10000}}\approx 0.00156$. Out of 162 simulations only 9 exceeded the resulting 95\% confidence interval which was true for fixed designs and re-calculation designs.

\begin{figure}[h!] 
\sbox0{\BLAsolid}\sbox1{\BLUsolid}\sbox2{\GREsolid}\sbox3{\REDsolid}
\sbox4{\BLUdash}\sbox5{\GREdash}\sbox6{\REDdash}\sbox7{\ORAsolid}\sbox8{\ORAdash}
\includegraphics[width=\textwidth]{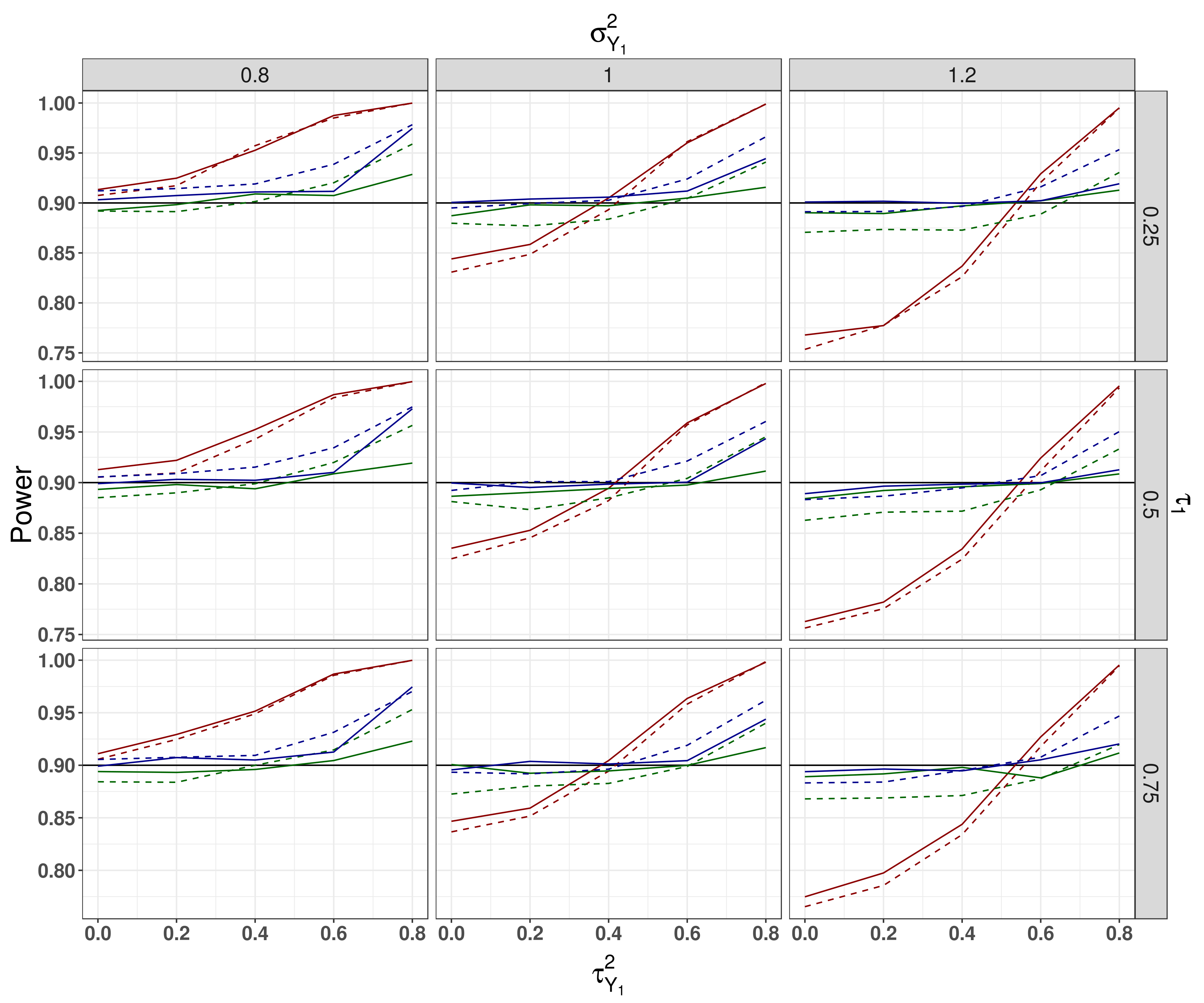} 
\caption{Power for a design testing composite populations $G_{I_1}=S_1$ and $G_{I_2}=S_1\cup S_2$. The power characteristics are plotted against varying values of true correlation ($\rho_{Y_1,X_1}^2$), true variance of subset $S_1$ ($\sigma_{Y_1}^2$) and varying prevalences $\tau_1$. Given a desired power of $90\%$ (\usebox0), the simulations results for treatment benefit $\beta_{11}=0.5$ are presented as follows: no IPS (\usebox3), IPS at $\nu=0.3$ (\usebox2) and IPS at $\nu=0.5$ (\usebox1). The results for a treatment benefit of $\beta_{11}=1$ are presented as follows: no IPS (\usebox6), IPS at $\nu=0.3$ (\usebox5) and an IPS at $\nu=0.5$ (\usebox4).)
}
\end{figure}

An improvement in the performance is seen when comparing the power of fixed designs ($\beta_{11}=0.5$:\usebox3 , $\beta_{11}=1$\usebox6) to designs which conducted an earlier ($\beta_{11}=0.5$\usebox2 , $\beta_{11}=1$:\usebox5) or later ($\beta_{11}=0.5$\usebox1 , $\beta_{11}=1$\usebox4) sample size re-calculation. See Figure 2. It stands out that in simulations where no IPS is conducted the power diverged widely from the desired power. However, in simulations where an IPS was conducted after $N_0\cdot 0.5$ or $N_0\cdot 0.3$ subjects were recruited, the power was met in most circumstances. An overpowering is observed in studies where high correlation between the outcome and the covariate exist. In those scenarios the required sample size to meet the desired power is rather low, so that the BSSR takes place at a point where already more than enough subjects have been recruited. This feature is also expressed in those simulations with an earlier IPS. Here, however, the overpowering is not as large, due to the earlier sample size re-calculation.\\
When comparing the simulations with earlier to those with later  re-calculation, we see a loss of  power by an average of 1\% to 2\%. This power loss becomes more severe the fewer subjects are present during the IPS. For scenarios where the IPS was conducted with only 13 subjects in $S_1$, the loss of power was 3-4\%. Therefore, the blinded sample size re-calculation should only be  carried out in the presence of at least 20 to 25 subjects in the smallest subset (see \citet{SANDVIK1996}).\\   \sbox0{\BLAsolid} \sbox4{\GREdot}\sbox5{\REDdot}
The panels of Figure 3 show the sample size when all parameters are known from the get go (\usebox0), the average sample sizes and the spread of the re-calculated sample sizes. Three points stand out. First, on average earlier BSSR (\usebox1) and later BSSR (\usebox2) yield the same average re-calculated sample size. Second, while the average sample sizes are the same, the spread of the sample size is always higher in earlier IPS (\usebox4) compared to a later IPS (\usebox5). And third, by comparing the re-calculated sample sizes with the optimal sample size, the average sample size was always higher than the optimal sample size. This amounted to 20 subjects too many when conducting an BSSR. This can be seen as the cost when estimating the parameters blindly. This increase of the average sample sizes is however independent of the required sample sizes as noted by \citet{Friede2001}.\\
 
\begin{figure}[H] 
\sbox0{\BLAsolid}\sbox1{\GREsolid}\sbox2{\BLUsolid}\sbox3{\ORAsolid}\sbox4{\GREdot}\sbox5{\BLUdot}\sbox6{\ORAdot}
\centering
\begin{minipage}[c]{.49\textwidth}
\centering a) \large{ $\beta_{11}=0.5$}
\includegraphics[width=\textwidth]{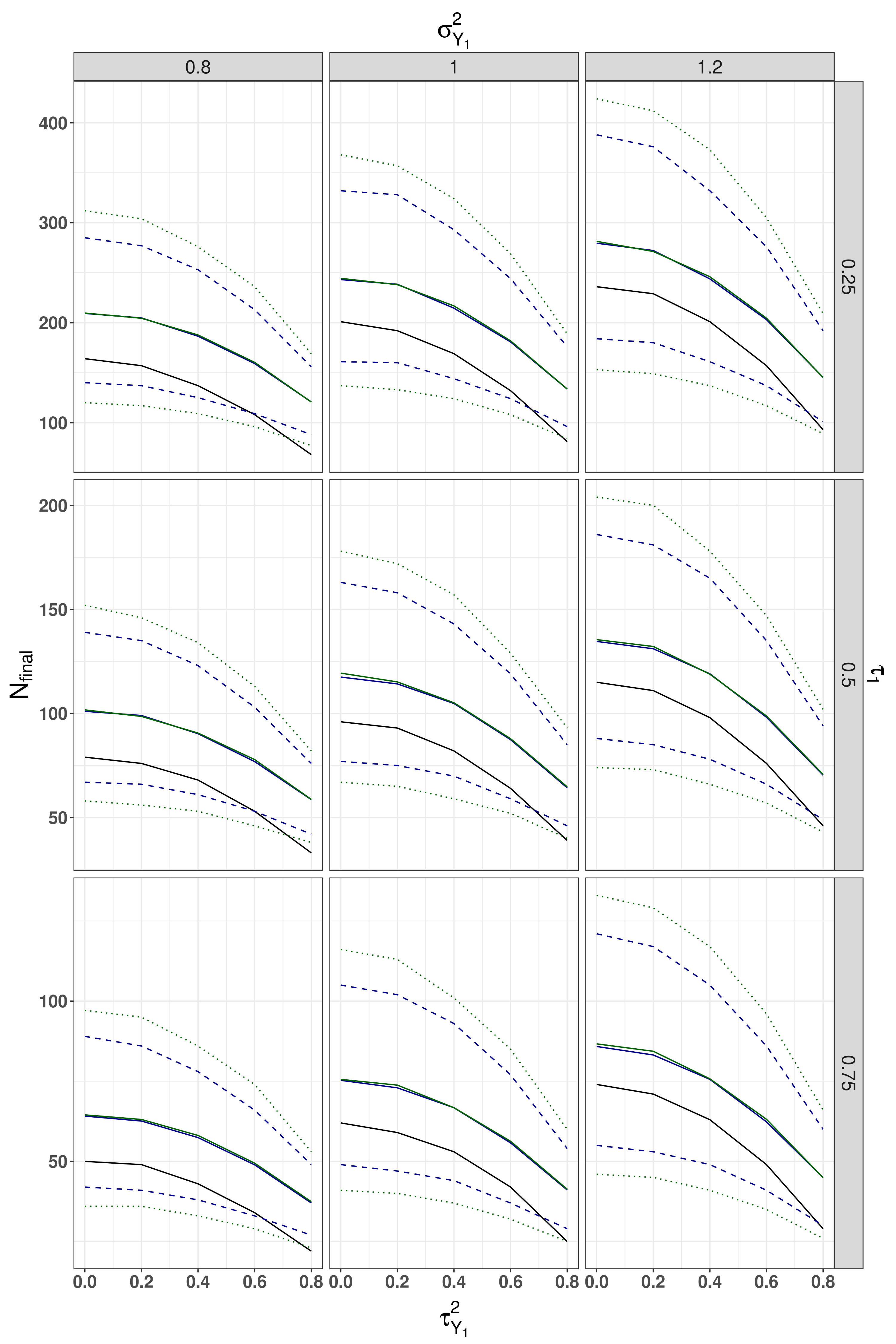} 
\end{minipage}
\begin{minipage}[c]{.49\textwidth}
\centering b) \large{ $\beta_{11}=1$}
\includegraphics[width=\textwidth]{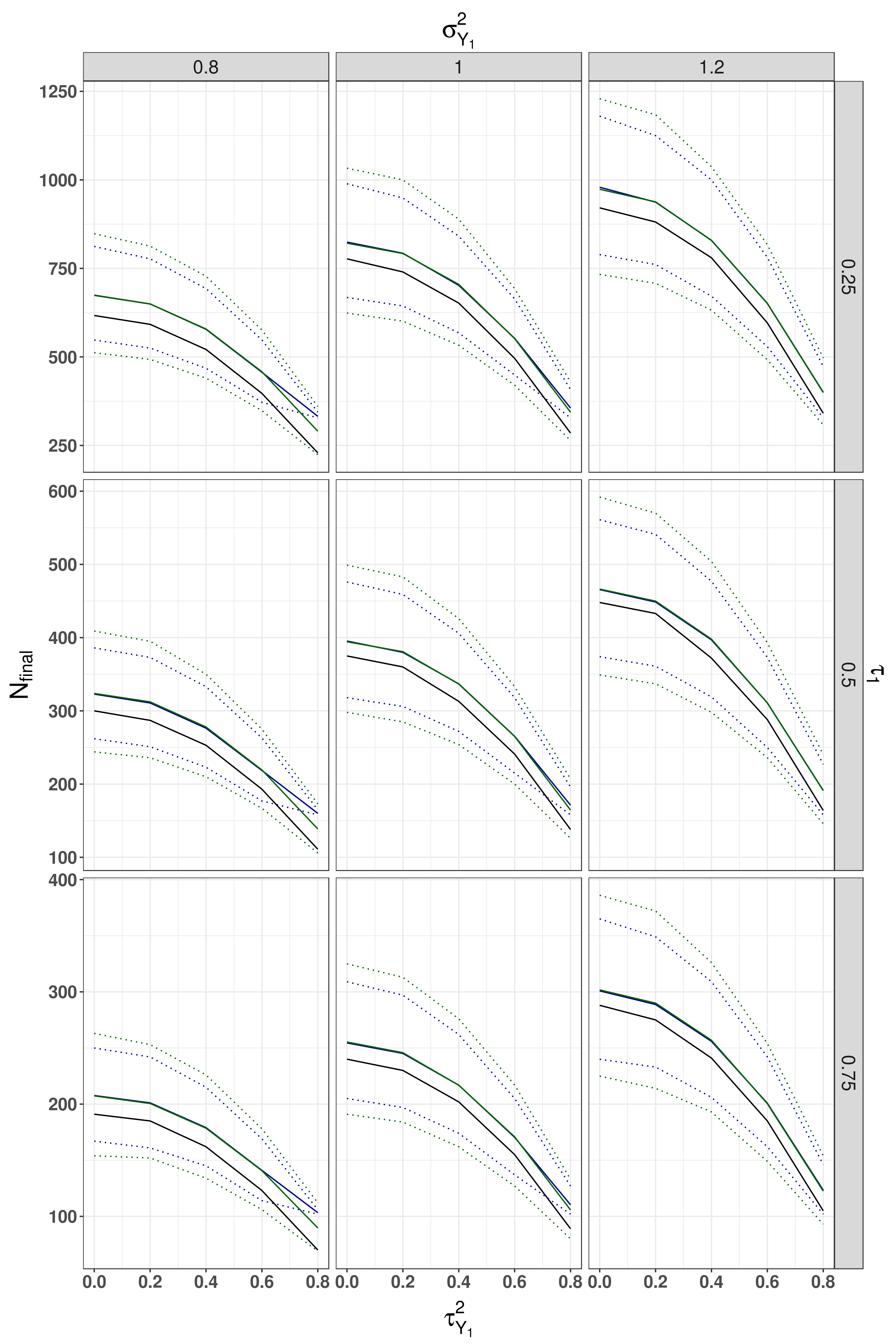} 
\end{minipage}
\caption{Calculated and re-calculated sample size for $\beta_{11}=0.5$ (a) and $\beta_{11}=1$ (b). The optimal sample size when using the unknown true values for each scenario are given by (\usebox0). The average re-calculated sample sizes for $\nu=0.3$ (\usebox1) and for $\nu=0.5$ (\usebox2). The spread using the 10\% and 90\% quantile sample sizes are highlighted by (\usebox4) and (\usebox5).} 
\end{figure}

\section{Discussion and Conclusion}

In this work, we considered the issue of testing treatment efficacy in a heterogeneous study population. We provided a simple approach to investigate a treatment effect in any composite population setting by testing treatments within pre-defined disjoint subsets first and re-assembling them to test treatment efficacy in composite populations afterwards. Therefore, we provided a simple testing strategy which controls the family-wise type I error rate in the strong sense for large and small sample sizes. While we considered normally distributed outcomes and any number of random baseline covariates our testing strategy is applicable to any $p$-value generating procedures which fulfil the \textit{p-clud} condition. We provided schemes to calculate the sample sizes for normally distributed outcomes and any number of covariates. This sample size calculation relies on correct a priori specifications of a number of nuisance parameters which increases the chance of misspecification. Hence, we include options for re-calculating the sample size in an internal pilot study.\\
Simulations showed that the target power was achieved if the initial parameter specifications were met, while it was not achieved under parameter misspecifications. The sample size re-calculation on the other hand maintained the desired power even in scenarios where the initial parameter assumptions were not met and the smallest subset had at least 20 subjects for the internal pilot study. The last part is in line with findings of \citet{SANDVIK1996} and \citet{Birkett1994}, who concluded that an internal pilot study should include 20 or more degrees of freedom. Contrasting the fixed design with the blinded sample size re-calculation design, an increased average sample size was noticed after re-calculation. This occurs due to an inflation of the residual variance when ignoring the treatment allocation. However, this inflation has no influence on the type I error rate. Also, many governmental agencies prefer blinded re-calculation because fewer sources for bias exist \citep{EMA1998,EMEA2007,FDA1,FDA2}. When comparing an earlier blinded sample size re-calculation and later blinded sample size re-calculation, a slight reduction in power was observed. This reduction is attributed to an increase in variation when conducting a blinded re-calculation with fewer subjects.\\
With regards to early sample size re-calculation, \citet{Zucker1999} addressed this issue for the simple setting of a two-sample t-test in one population  by multiplying the re-calculated sample size by inflation factor $\left(t_{\alpha,\eta_1}+t_{\beta,\eta_1}\right)^2/\left(\Phi^{-1}(\alpha)+\Phi^{-1}(\beta)\right)$. Here, $t_{\alpha,\eta_1}$ and $t_{\beta,\eta_1}$ denote the t-quantiles given $\alpha, \beta$ while $\eta_1$ denote the degrees of freedom during IPS. \citet{Placzek2017} extended that idea for a multiple nested subpopulation setting. In their work, the intersection hypothesis is tested using a multivariate t-distribution with degrees of freedom according to either the full population or the smallest subgroup. To account for uncertainties during the IPS, degrees of freedom according to the full population or the smallest subgroup at the IPS are used. Both procedures improve the results of the power analysis at the cost of higher sample sizes. An alternative approach to the inflation factor was discussed by \citet{Kieser2000a} and \citet{Friede2006}, where instead of using the point estimate of the residual variance, an upper confidence bound is used. \\
A future route to extend the methods provided here is to incorporate the adaptive enrichment design \citep{Wang2007,Brannath2009,Jenkins2011a,Friede2012,Placzek2019}. Adaptive trials are made up of two or more stages where the efficacy of a treatment is evaluated at interim analyses between stages. During the first stage subjects from the whole population are recruited but a change of the recruitment plan can be initiated during an interim analysis when this change was pre-defined at the start of the study.

\noindent {\bf{Conflict of Interest}}

\noindent {\it{The authors have declared no conflict of interest.}}
\newpage
\section*{Appendix}
\subsection*{Simulations for planned power = 80\%}

\begin{table}[h]
\caption{Comparing estimated probabilities of rejection of the intersection hypothesis $H_{0G_{\mathcal{I}_1}\cap G_{\mathcal{I}_2} }$ for fixed design and designs with blinded sample size re-calculation. The true and assumed values can be found in Table 1. Let $N_0$ denote the initially calculated total sample size using assumed variance and correlation but true prevalences. The predicted 95\%-confidence interval is calculated as [0.02194;0.02806] given a nominal level of $\alpha=0.025$. Values which fall out of the 95\%-confidence interval are highlighted. Out of the 162 simulations, scenarios 9 where out of bounds.} 
\centering
\begin{tabular}{ccccccccccc}
\toprule
  & &  \multicolumn{1}{l}{}& \multicolumn{4}{l}{$\mathbf{\beta_{11}=0.5}$} & \multicolumn{4}{l}{$\mathbf{\beta_{11}=1}$}\\
$\tau_1$ & $\sigma_{Y_1}^2$ &\multicolumn{1}{l}{ $\rho_{Y_1,X_1}^2=\rho_{Y_2,X_2}^2$} & $N_0$ &  \shortstack{FWER\\ no IPS} &  \shortstack{FWER\\ $\nu=0.3$} & \shortstack{FWER\\ $\nu=0.5$} &\multicolumn{1}{l}{$N_0$} &  \shortstack{FWER\\ no IPS} &  \shortstack{FWER\\ $\nu=0.3$} & \shortstack{FWER\\ $\nu=0.5$}  \\
\midrule
\rowcolor{gray!6}  0.25 & 0.8 & 0.0 & 493 & 0.0246 & 0.0249 & 0.0253 & 133 & 0.0247 & 0.0245 & 0.0234\\
\rowcolor{gray!6}       &     & 0.4 &     & 0.0247 & 0.0257 & 0.0234 &     & 0.0247 & 0.0258 & 0.0263\\
\rowcolor{gray!6}       &     & 0.8 &     & 0.0254 & 0.0245 & 0.0251 &     & 0.0223 & 0.0231 & 0.0246\\
\rowcolor{gray!6}       & 1   & 0.0 &     & 0.0257 & 0.0267 & 0.0237 &     & 0.0271 & 0.0239 & 0.0244\\
\rowcolor{gray!6}       &     & 0.4 &     & 0.0252 & 0.0264 & 0.0256 &     & 0.0230 & 0.0267 & 0.0270\\
\rowcolor{gray!6}       &     & 0.8 &     & 0.0242 & 0.0263 & 0.0254 &     & 0.0240 & 0.0244 & 0.0261\\
\rowcolor{gray!6}       & 1.2 & 0.0 &     & 0.0258 & 0.0264 & 0.0249 &     & 0.0238 & 0.0246 & 0.0235\\
\rowcolor{gray!6}       &     & 0.4 &     & 0.0277 & 0.0235 & 0.0266 &     & 0.0249 & 0.0238 & 0.0268\\
\rowcolor{gray!6}       &     & 0.8 &     & 0.0247 & 0.0242 & 0.0274 &     & 0.0238 & 0.0264 & 0.0257\\
\addlinespace
0.50 & 0.8 & 0.0 & 239 & 0.0247 & 0.0222 & 0.0241 & 64 & 0.0223 & 0.0253 & 0.0238\\
     &     & 0.4 &     & 0.0249 & 0.0244 & 0.0251 &    & 0.0251 & 0.0234 & 0.0238\\
     &     & 0.8 &     & 0.0249 & 0.0237 & 0.0250 &    & 0.0261 & 0.0255 & 0.0242\\
     & 1   & 0.0 &     & 0.0270 & 0.0238 & 0.0252 &    & 0.0286 & 0.0262 & 0.0245\\
     &     & 0.4 &     & 0.0226 & 0.0250 & 0.0247 &    & 0.0261 & 0.0232 & 0.0260\\
     &     & 0.8 &     & 0.0250 & 0.0227 & 0.0249 &    & 0.0234 & 0.0238 & 0.0234\\
     & 1.2 & 0.0 &     & 0.0228 & 0.0239 & 0.0249 &    & 0.0236 & 0.0228 & 0.0251\\
     &     & 0.4 &     & 0.0236 & 0.0251 & 0.0240 &    & 0.0259 & 0.0270 & 0.0247\\
     &     & 0.8 &     & 0.0266 & 0.0255 & 0.0260 &    & 0.0249 & 0.0246 & 0.0265\\
\addlinespace
\rowcolor{gray!6}  0.75 & 0.8 & 0.0 & 151 & 0.0262 & 0.0249 & \textbf{0.0281} & 41 & 0.0226 & 0.0246 & 0.0278\\
\rowcolor{gray!6}       &     & 0.4 &     & 0.0261 & 0.0251 & 0.0227 &    & 0.0253 & 0.0233 & 0.0275\\
\rowcolor{gray!6}       &     & 0.8 &     & 0.0236 & 0.0273 & 0.0277 &    & 0.0238 & 0.0277 & 0.0251\\
\rowcolor{gray!6}       & 1   & 0.0 &     & 0.0237 & 0.0263 & 0.0238 &    & 0.0245 & 0.0246 & 0.0239\\
\rowcolor{gray!6}       &     & 0.4 &     & 0.0258 & 0.0259 & 0.0251 &    & 0.0254 & 0.0258 & 0.0254\\
\rowcolor{gray!6}       &     & 0.8 &     & 0.0239 & 0.0244 & 0.0268 &    & 0.0256 & 0.0258 & \textbf{0.0296}\\
\rowcolor{gray!6}       & 1.2 & 0.0 &     & 0.0260 & 0.0235 & 0.0267 &    & 0.0269 & 0.0255 & 0.0231\\
\rowcolor{gray!6}       &     & 0.4 &     & 0.0224 & 0.0231 & 0.0227 &    & 0.0242 & 0.0234 & 0.0244\\
\rowcolor{gray!6}       &     & 0.8 &     & 0.0255 & 0.0234 & 0.0254 &    & 0.0261 & 0.0248 &\textbf{0.0219}\\
\bottomrule
\end{tabular}
\end{table}

\begin{figure}[H] 
\sbox0{\BLAsolid}\sbox1{\BLUsolid}\sbox2{\GREsolid}\sbox3{\REDsolid}
\sbox4{\BLUdash}\sbox5{\GREdash}\sbox6{\REDdash}\sbox7{\ORAsolid}\sbox8{\ORAdash}
\includegraphics[width=\textwidth]{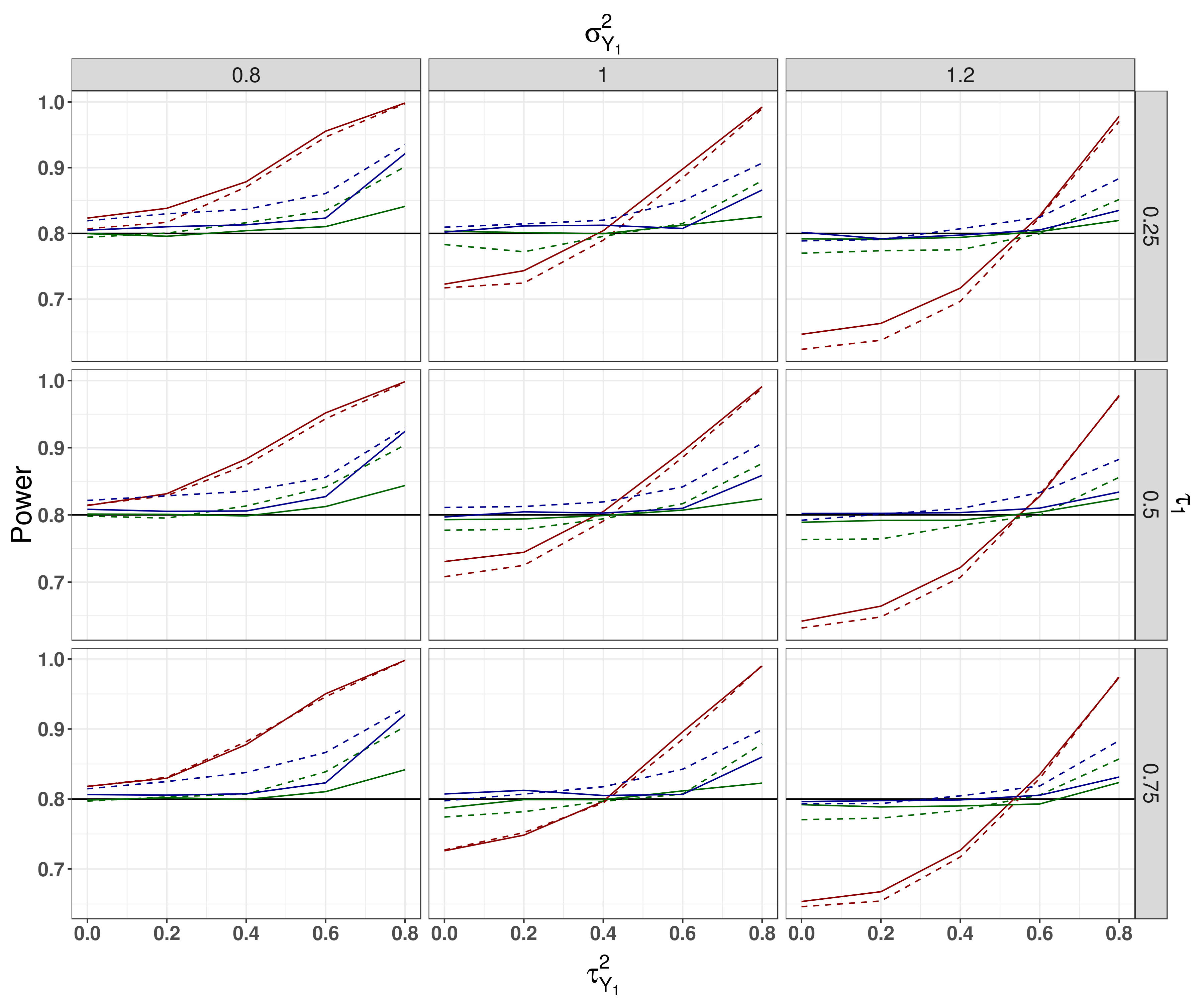} 
\caption*{Power for a design testing composite populations $G_{I_1}=S_1$ and $G_{I_2}=S_1\cup S_2$. The power characteristics are plotted against varying values of true correlation ($\rho_{Y_1,X_1}^2$), true variance of subset $S_1$ ($\sigma_{Y_1}^2$) and varying prevalences $\tau_1$. Given a desired power of $80\%$ (\usebox0), the simulations results for treatment benefit $\beta_{11}=0.5$ are presented as follows: no IPS (\usebox3), IPS at $\nu=0.3$ (\usebox2) and IPS at $\nu=0.5$ (\usebox1). The results for a treatment benefit of $\beta_{11}=1$ are presented as follows: no IPS (\usebox6), IPS at $\nu=0.3$ (\usebox5) and an IPS at $\nu=0.5$ (\usebox4).)}
\end{figure}

\begin{figure}[H] 
\sbox0{\BLAsolid}\sbox1{\GREsolid}\sbox2{\BLUsolid}\sbox3{\ORAsolid}\sbox4{\GREdot}\sbox5{\BLUdot}\sbox6{\ORAdot}
\centering
\begin{minipage}[c]{.49\textwidth}
\centering a) \large{ $\beta_{11}=0.5$}
\includegraphics[width=\textwidth]{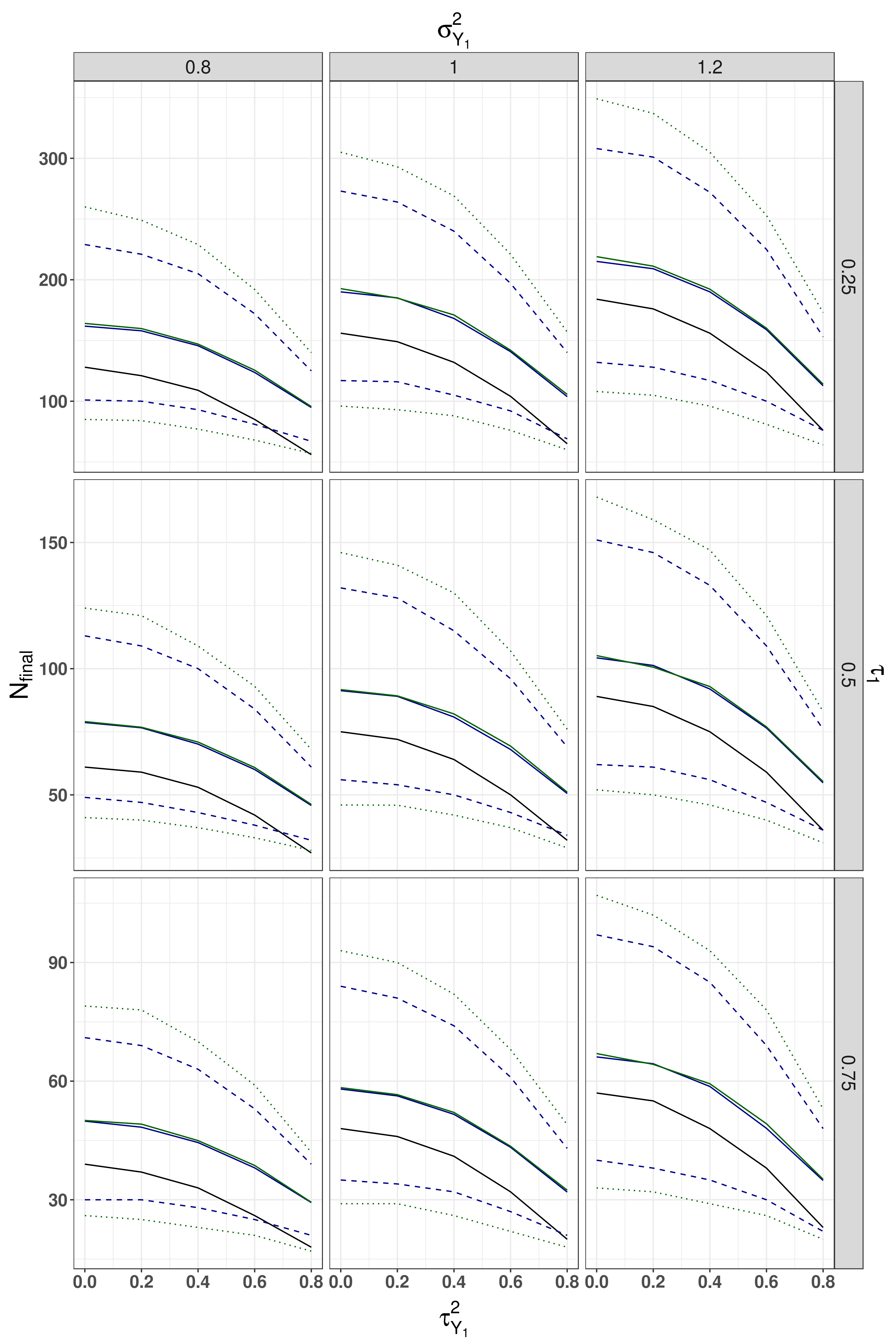} 
\end{minipage}
\begin{minipage}[c]{.49\textwidth}
\centering b) \large{ $\beta_{11}=1$}
\includegraphics[width=\textwidth]{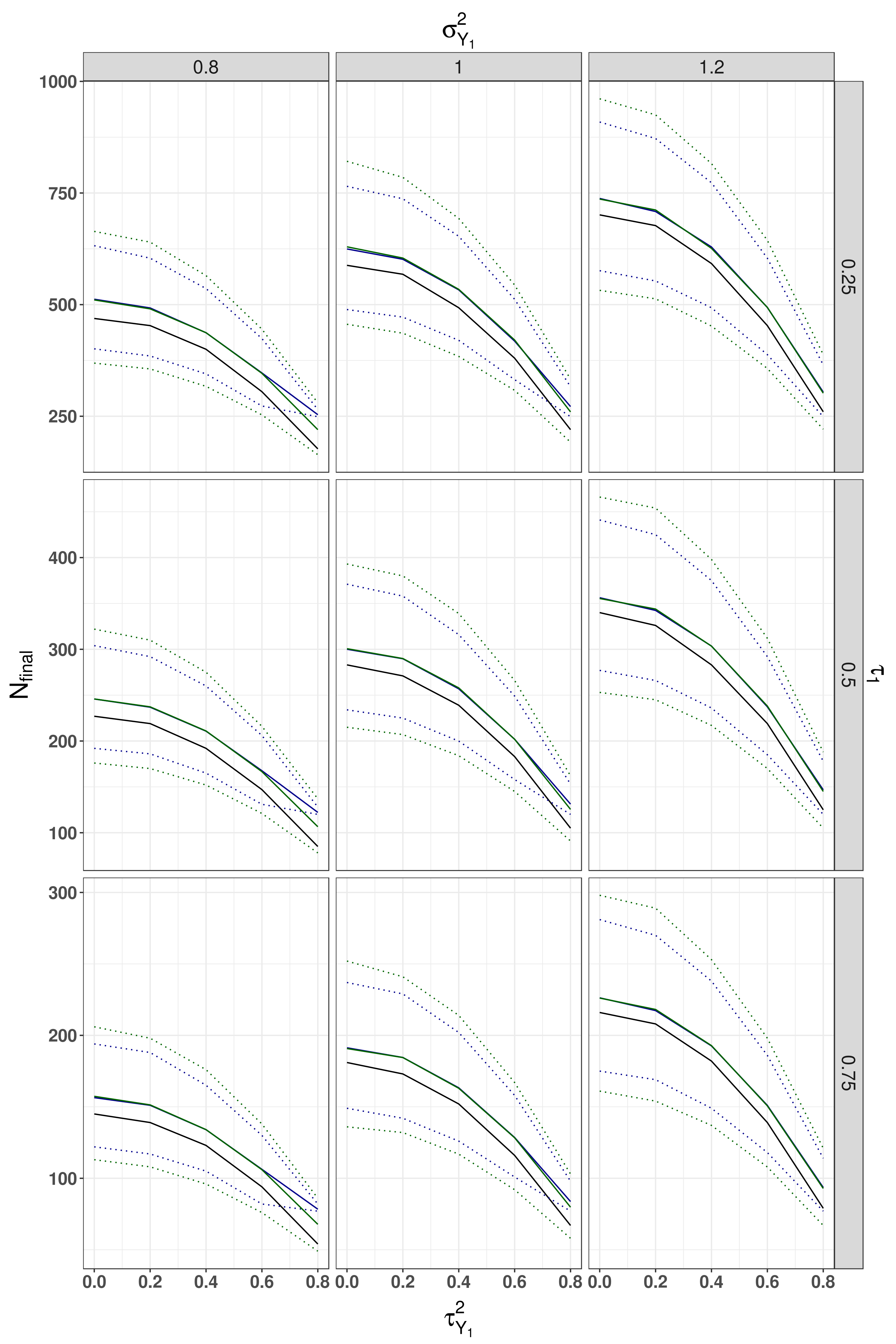} 
\end{minipage}
\caption*{Calculated and re-calculated sample size for $\beta_{11}=0.5$ (a) and $\beta_{11}=1$ (b). The optimal sample size when using the unknown true values for each scenario are given by (\usebox0). The average re-calculated sample sizes for $\nu=0.3$ (\usebox1) and for $\nu=0.5$ (\usebox2). The spread using the 10\% and 90\% quantile sample sizes are highlighted by (\usebox4) and (\usebox5).} 
\end{figure}
\newpage
\section*{}
{
\bibliographystyle{bimj}
\bibliography{literatur}
}

\phantom{aaaa}
\end{document}